\documentclass{llncs}

\usepackage{cite}

\ifx\pdfoutput\undefined
\usepackage{graphicx}
\else
\usepackage[pdftex]{graphicx}
\fi

\usepackage{psfrag}
\usepackage{subfigure}
\usepackage{url}
\usepackage{stfloats}

\usepackage{amsmath}

\usepackage{subfigure}

\usepackage{algorithm}

\newcommand{\eSTab}{\hspace{0.5cm}}
\newcommand{\eTab}{\hspace{0.8cm}}

\begin{document}

\title{Construction of Small Worlds \\
in the Physical Topology of Wireless Networks}

\titlerunning{Small Worlds}

\author{Fengfeng~Zhou \and
        Guoliang~Chen \and
        Yinlong~Xu
           \thanks{This work was supported by National Science Foundation of China [No.60173048].}
        }%

\institute{Department of Computer Science \\
University of Science and Technology of China, 230027 \\
Hefei, Anhui, P.R.China \\
hunterz@mail.ustc.edu.cn}

\maketitle

\begin{abstract}
The concept of small worlds is introduced into the physical topology of wireless networks in this work.
A.~Helmy provided two construction schemes of small worlds for the wireless networks,
link rewiring and link addition, but he mainly focused
on the virtual topology. Based on the broadcasting nature of the radio transmission, we propose a
construction scheme of small worlds for the physical topology of
Multiple-Input Multiple-Output (MIMO) wireless networks.
Besides the topology-related topics, we also evaluate the reduction of the power required by a request.


\end{abstract}

\section{Introduction}

The small world phenomenon was first discussed by S.~Milgram
\textit{et al.} \cite{first_1,first_2} (also known as {\em six
degrees of separation} \cite{first_3}). D.J.~Watts \textit{et al.}
considered it in some real world situations, such as the
electrical power grids, the epidemical models of infectious
diseases, and the collaboration relations of actors, etc
\cite{book_1,nature_1}, which were called {\em small world
networks}. Much more research work has been stimulated in the
literature
\cite{nature_2,nature_3,small_world_1,small_world_2,small_world_3,small_world_4,small_world_5,small_world_6}.

The regular networks have large {\em clustering coefficient}
\footnote{fraction of nodes' neighbors that are also neighbors of
each other} and large {\em characteristic path hop}
\footnote{average hops of the shortest paths between nodes}, while
the random networks with the same size and average node degree
have much smaller clustering coefficient and characteristic path
hop. With introducing some "short-cuts" into the regular networks
by rewiring each edge with probability $p$, D.J.~Watts \textit{et
al.} constructed the small world networks and observed that the
characteristic path hop decreases dramatically as $p$ increases,
but the clustering coefficient decreases slowly.

The multi-hop wireless communication networks own high clustering due to their broadcasting
nature, which leads to large characteristic path hop compared to the random networks. Ahmed~Helmy
\textit{et al.} proposed two construction schemes of small worlds in such
networks: link rewiring and link addition \cite{small_world_6}, and studied the concept
in the virtual topology \cite{small_world_7,small_world_8}.

Two virtual "shot-cut" links may share a physical link between two nodes, which makes them
interfering with each other. Due to the broadcasting nature of the wireless networks, there
may be many such interferences for the above construction schemes of small worlds in the
virtual topology. In order to throw off these shortcomings, and
still retain the short characteristic path hop, we apply the small world concept into
the physical topology of multi-hop wireless networks in a
multiple-input multiple-output (MIMO) manner. With $k$ pairs of transmitting and receiving antennas at each
node, the bandwidth of such MIMO wireless networks is equivalent to the sum of the capacity of $k$ parallel
single-input single-output (SISO) channels \cite{MIMO}. So the radio spectrum is divided into $k$
channels with equal bandwidth. One of the channels acts as the normal data channel, called the {\em
normal channel}, while the others are dedicated to the "short-cut" communications, called the {\em
short-cut channels}. We construct the short-cuts over the short-cut channels, and evaluate some practical
objectives in such MIMO wireless networks.

The remainder of this paper is organized as follows. Section \ref{Sec_2} gives the problem model and some
definitions. A construction scheme of the short-cut channels
 is presented in Section \ref{Sec_3}. The performance evaluation with some
numerical results are given in Section \ref{Sec_4}. Section \ref{Sec_5} concludes.

\section{Preliminary}\label{Sec_2}

Since there must be some ACK packets between the two terminals of a request in most wireless protocols, we use an
undirected weighted graph $G=(V,E)$ to represent the MIMO wireless network.

The power that node $a$ needs to transmit at the radio range $R_a$ is proportional to $R_{a}^\alpha$, where
the {\em power constant} $\alpha$ is a
parameter ranging between 1 and 4,
depending on the communication environment. Without loss of generality, we set the normalizing constant to 1.
Let the transmitting ranges of node $a$ and $b$ be $R_a$ and $R_b$ respectively, and let
the distance between $a$ and $b$ be $R(a,b)$. If $R_a \ge R(a, b)$ and
$R_b \ge R(a, b)$, there exists an edge $(a, b) \in E$, and the power that the edge needs is:
$p_{(a,b)}= \max{\{R_a^\alpha, R_b^\alpha\}}$, which is denoted as the weight of the edge.

\begin{definition}[Characteristic Path Hop and Path Length]
In the MIMO wireless network, the number of edges along the shortest path between two nodes $a$ and $b$ is
called the {\em path hop} between $a$ and $b$,
denoted as $H(a,b)$, and the sum of the weight of these edges is called the
{\em path length} between $a$ and
$b$, denoted as $L(a,b)$. The {\em characteristic path hop} and {\em characteristic path length} of
the network are defined as the average path hop and path length over all connected pairs of nodes respectively.
\end{definition}

The characteristic path hop \footnote{This concept is known as {\em characteristic path length} in
most previous papers. In this paper, we use {\em characteristic path hop} instead, and
redefine {\em path length} to represent the power requirement of
a request.}
shows the separation of a network. There is an interesting discovery that the
characteristic path hop of most real world networks is relatively small, even when these kinds of networks
have many fewer edges than a typical globally coupled network with the same number of nodes. This observation
stimulates the consideration of applying the small world concept into the communication networks.
The characteristic path length represents the average power that a request needs.

\begin{definition}[Clustering Coefficient]
In the MIMO wireless network, node $a \in V$ has $k$ neighbors. The ratio between the number $E_a$ of edges
actually existing among node $a$'s neighbors and the total possible number $k(k-1)/2$ is called the {\em
clustering coefficient} $C_a$ of node $a$. The {\em clustering coefficient} $C$ of the network is the average
of $C_a$ over all the nodes in $V$.
\end{definition}

In this work, we focus on the {\em Media Access Control} (MAC) protocols in the physical topology
of MIMO wireless networks, 
in which no node mobility is considered, and the channel condition remains unchanged.

\begin{figure}
\centering
\subfigure[Random]{\label{FourNDist.Random} \includegraphics[angle=-90, width=4.2cm,totalheight=4.2cm]{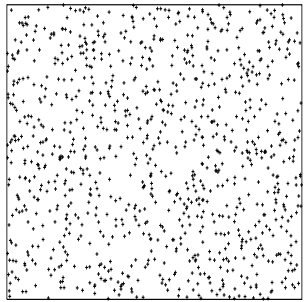}}
\subfigure[Normal]{\label{FourNDist.Normal} \includegraphics[angle=-90,width=4.2cm, totalheight=4.2cm]{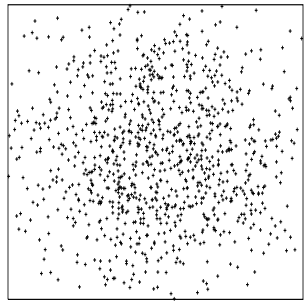}}
\subfigure[Skewed]{\label{FourNDist.Skewed}\includegraphics[angle=-90, width=4.2cm, totalheight=4.2cm]{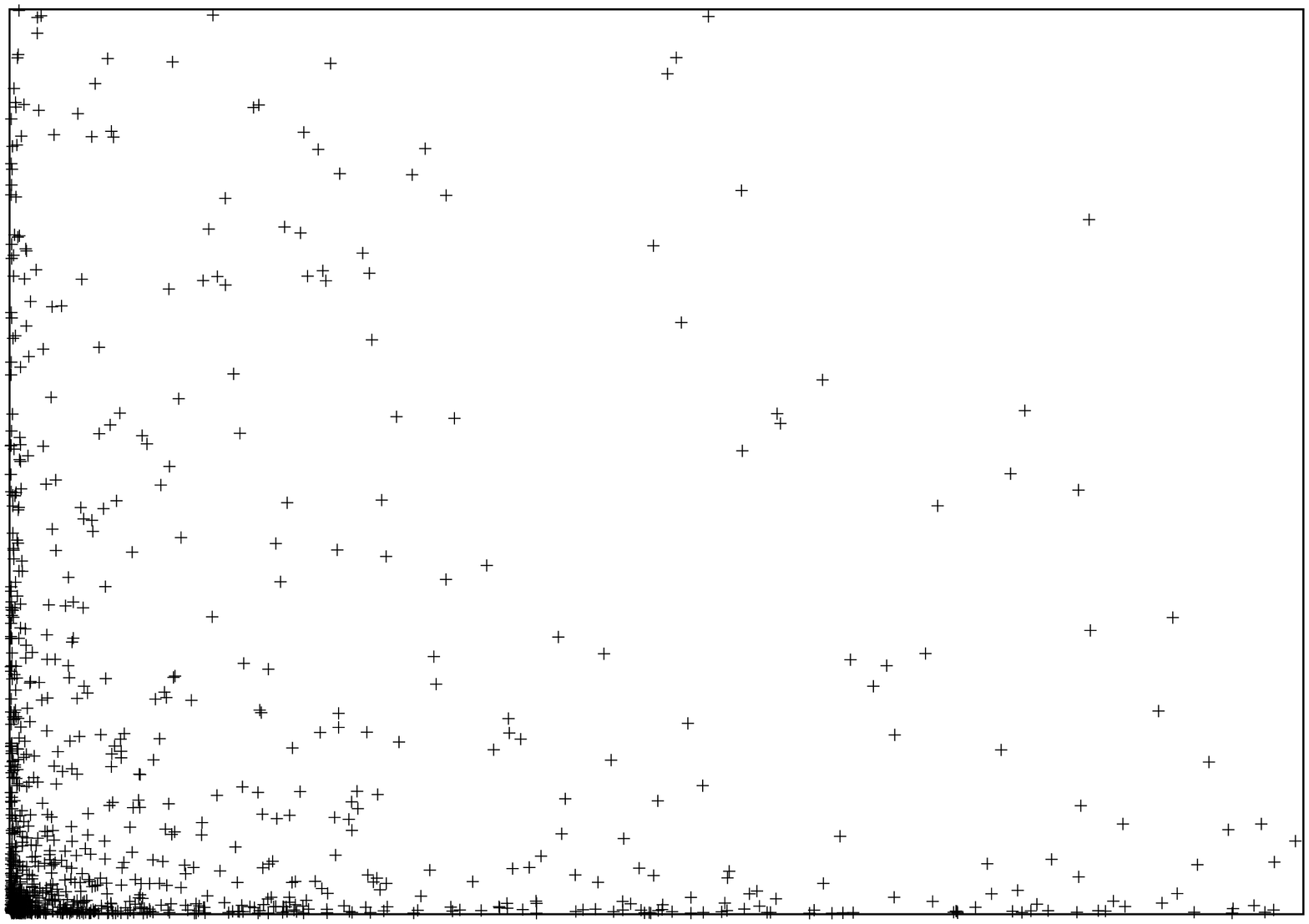}}
\subfigure[Grid]{\label{FourNDist.Grid}\includegraphics[angle=-90, width=4.2cm, totalheight=4.2cm]{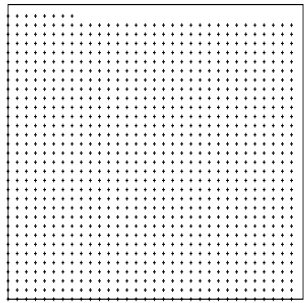}}
\caption{Four node distributions with 1000 nodes over a $1 km \times 1 km$ area}
\label{Fig_FourNDist}
\end{figure}

\section{Construction of Small worlds}\label{Sec_3}

Without loss of generality, we conduct our simulations with 1000 nodes over a $1 km \times 1 km$ area. We
investigate four node distributions, including random, normal, skewed, and grid, and several broadcasting
ranges to represent different network layouts.

Ahmed~Helmy \cite{small_world_6} proposed two short-cut construction schemes, link rewiring and
link addition, for the virtual topology of wireless networks. For link rewiring, a node is randomly chosen, and
then a link to one of its neighbors is removed and relinked to a random node. For link addition, a
pair of nodes are randomly chosen, and connected with a link.

In the physical topology of wireless networks without mobility, the established links may not be removed or rewired,
while the addition of a "long-distance" link may introduce a large amount of interference into the wireless
networks.
So we construct the short-cuts over the short-cut channels in the MIMO wireless network. According to
I.E.~Telatar's work \cite{MIMO}, we can reasonably anticipate the performance improvement over
the widely-deployed wireless networks with SISO topologies, such as the
IEEE 802.11 protocols.

Below we show how to construct the short-cuts in the above MIMO wireless networks.

The MIMO wireless network is denoted as an undirected weighted graph $G=(V,E)$. The location of nodes
in $V$ is decided according to one of the four node distributions: random, normal, skewed, and grid, as
illustrated in Fig. \ref{Fig_FourNDist}. There are $(k+1)$ communication channels: one normal channel $C_0$
and $k$ short-cut channels $C_1$, $C_2$, $\cdots$, $C_k$ $(k \ge 0)$. For a pair of nodes $a$ and $b$ with
radio range $R_a$ and $R_b$ respectively over the channel $C_i$ $(0\le i \le k)$, if $R(a,b) \le R_a$, we
say that node $b$ is {\em covered} by node $a$ over the channel $C_i$. If node $a$ and $b$ are covered by
each other over the channel $C_i$, there exists an edge $(a,b)$ in the set of edges $E_i$ $(0\le i \le k)$.
For node $c \in V$, if $c$ is covered by $a$ or $b$, we say that node $c$ is {\em covered} by the edge $(a,
b)$. So we have: $E=E_0 \cup E_1 \cup \cdots \cup E_k$.

As in most wireless protocols, the radio range over the normal channel $C_0$ is a fixed constant $R_0$. So
there exists an edge $(a,b) \in E_0$, if the distance between node $a$ and $b$ satisfies: $R(a,b) \le R_0$.
Such an edge is called a {\em normal edge}, of which the weight is $R_0^\alpha$.


\floatname{algorithm}{Procedure}
\begin{algorithm}
\caption{BuildSCChannel} \label{Proc_BuildSCChannel}

\eSTab \begin{quote} {\bf Input:}

The nodes: $Nodes$,\\
the edges in the established channels: $Edges$,\\
and the ratio over $R_0$  of the upper bound of
the radio range: $RadiiRatio$.

\vspace{0.2cm}

$SC\_Edges{} = Nil$;

{\em do}

\{

\eTab Randomly choose a pair of nodes $a$ and $b$ satisfying:
\eTab \begin{quote}
\begin{enumerate}
\item Edge $(a,b)$ does not exist in $Edges$;
\item in $SC\_Edges \cup \{(a,b)\}$, there doesn't exist an edge whose terminals are covered by other edges;
\item the distance between $a$ and $b$ $R(a,b)$ satisfying: $R_0 < R(a,b) \le R_0 \times RadiiRatio$.
\end{enumerate}
\end{quote}

\eTab $Edges = Edges \cup \{(a,b)\}$;

\eTab $SC\_Edges = SC\_Edges \cup \{(a,b)\}$;

\} {\em while} (There exists such an edge);

{\em return} $SC\_Edges$;
\end{quote}
\end{algorithm}

Different from the previous small world networks, in the MIMO wireless networks, some edges may interfere
with each other, and cannot act as transmitting links simultaneously. To avoid such interference, which will
greatly increase the complexity of the wireless routing protocol, we construct a short-cut channel with edges
whose terminals are not covered by other edges over the current channel, as described
in the procedure {\em BuildSCChannel}. To increase the minimum lifetime of nodes in $V$, we prevent
the construction of parallel links.
We also limit the radio range of the short-cuts by the upper bound $R_0 \times RadiiRatio$.

We construct the MIMO wireless networks with one {\em compound channel}, which consists of one normal
channel and several short-cut channels. It can be reasonably anticipated that such networks will outperform
the widely-deployed wireless networks with SISO topologies, as indicated in the following experiments on
some topology-related topics and the power efficiency. Observing from the above construction scheme,
in the MIMO wireless network with one compound
channel, the short-cut edges can act as transmitting links simultaneously, and they don't interfere with the
normal edges, too.
Due to the broadcasting nature of the wireless networks, and the distributed detection of interference,
the short-cuts can be constructed distributedly, which will improve the system performance of such MIMO wireless
networks greatly.

\begin{table}
\renewcommand{\arraystretch}{1.3}
\caption{The clustering coefficient, characteristic path hop, maximum path hop,
characteristic path length, and
maximum path length for the investigated topologies without short-cut channels ($k=0$).}
\label{Tbl_Topology}
\centering

\begin{tabular}{|c|c|c|c|c|c|c|c|}
\hline
{\bf Topology} & {\bf Range ($m$)} & {\bf Links} & {\bf $C(0)$} & {\bf $H(0)$} & {\bf $M(0)$} & {\bf $L(0)$} & {\bf $m(0)$} \\
\hline
Random Graph \cite{small_world_6} & - & - & 0.009 & 3.3 & 5 & - & - \\
\hline
Random-40 & 40 & 2305 & 0.550 & 24.956 & 57 & 998.253 & 2280 \\
\hline
Random-50 & 50 & 3645 & 0.576 & 15.564 & 39 & 778.208 & 1950 \\
\hline
Random-60 & 60 & 5265 & 0.589 & 11.907 & 30 & 714.414 & 1800 \\
\hline
Normal-60 & 60 & 8837 & 0.582 & 7.993 & 26 & 479.604 & 1560 \\
\hline
Skewed-50 & 50 & 70752 & 0.729 & 7.585 & 45 & 379.235 & 2250 \\
\hline
Grid-35 & 35 & 1936 & 0.000 & 21.121 & 62 & 739.222 & 2170 \\
\hline
Grid-60 & 60 & 3811 & 0.451 & 14.783 & 31 & 886.986 & 1860 \\
\hline
\end{tabular}
\end{table}

\begin{table}
\renewcommand{\arraystretch}{1.3}
\caption{Edges added in the short-cut channels.
(The number of edges over the normal channel is $|E_0|$.)}
\label{Tbl_SCEdge}
\centering
\begin{tabular}{c}
(a) Random-40 ($|E_0|=2520$) \\
    \begin{tabular}{|c|c|c|c|c|c|c|c|c|c|}
    \hline
    {\bf $i$} & 1 & 2 & 3 & 4 & 5 & 6 & 7 & 8 & 9 \\
    \hline
    {\bf $|E_i|$}  & 29 & 32 & 44 & 24 & 44 & 21 & 37 & 35 & 38 \\
    \hline
    {\bf $|SC_i|/|E_0|$ (\%)}  & 1.15 & 2.42 & 4.17 & 5.12 & 6.87 & 7.70 & 9.17 & 10.56 & 12.06 \\
    \hline
    \end{tabular}\\ \\
(b) Random-50 ($|E_0|=3888$) \\
    \begin{tabular}{|c|c|c|c|c|c|c|c|c|c|}
    \hline
    {\bf $i$} & 1 & 2 & 3 & 4 & 5 & 6 & 7 & 8 & 9 \\
    \hline
    {\bf $|E_i|$}  & 23 & 17 & 22 & 23 & 21 & 27 & 24 & 26 & 20 \\
    \hline
    {\bf $|SC_i|/|E_0|$ (\%)}  & 0.59 & 1.03 & 1.59 & 2.18 & 2.73 & 3.42 & 4.04 & 4.71 & 5.22 \\
    \hline
    \end{tabular}\\ \\
(c) Random-60 ($|E_0|=5505$) \\
    \begin{tabular}{|c|c|c|c|c|c|c|c|c|c|}
    \hline
    {\bf $i$} & 1 & 2 & 3 & 4 & 5 & 6 & 7 & 8 & 9 \\
    \hline
    {\bf $|E_i|$}  & 15 & 23 & 15 & 15 & 21 & 11 & 23 & 18 & 27 \\
    \hline
    {\bf $|SC_i|/|E_0|$ (\%)}  & 0.27 & 0.69 & 0.96 & 1.24 & 1.62 & 1.82 & 2.23 & 2.56 & 3.05 \\
    \hline
    \end{tabular}
\end{tabular}
\end{table}

\section{Numerical Results and Discussion}\label{Sec_4}

\psfrag{C_Str}{\LARGE{$C(k)/C(0)$: }}
\psfrag{H_Str}{\LARGE{$H(k)/H(0)$: }}
\psfrag{M_Str}{\LARGE{$M(k)/M(0)$: }}
\psfrag{L_Str}{\LARGE{$L(k)/L(0)$: }}
\psfrag{m_Str}{\LARGE{$m(k)/m(0)$: }}
\psfrag{k}{\LARGE{$k$}}

\begin{figure}
\centering
\subfigure[Random-40]{\label{PerfImp.40} \includegraphics[angle=-90, width=5.6cm,totalheight=3.7cm]{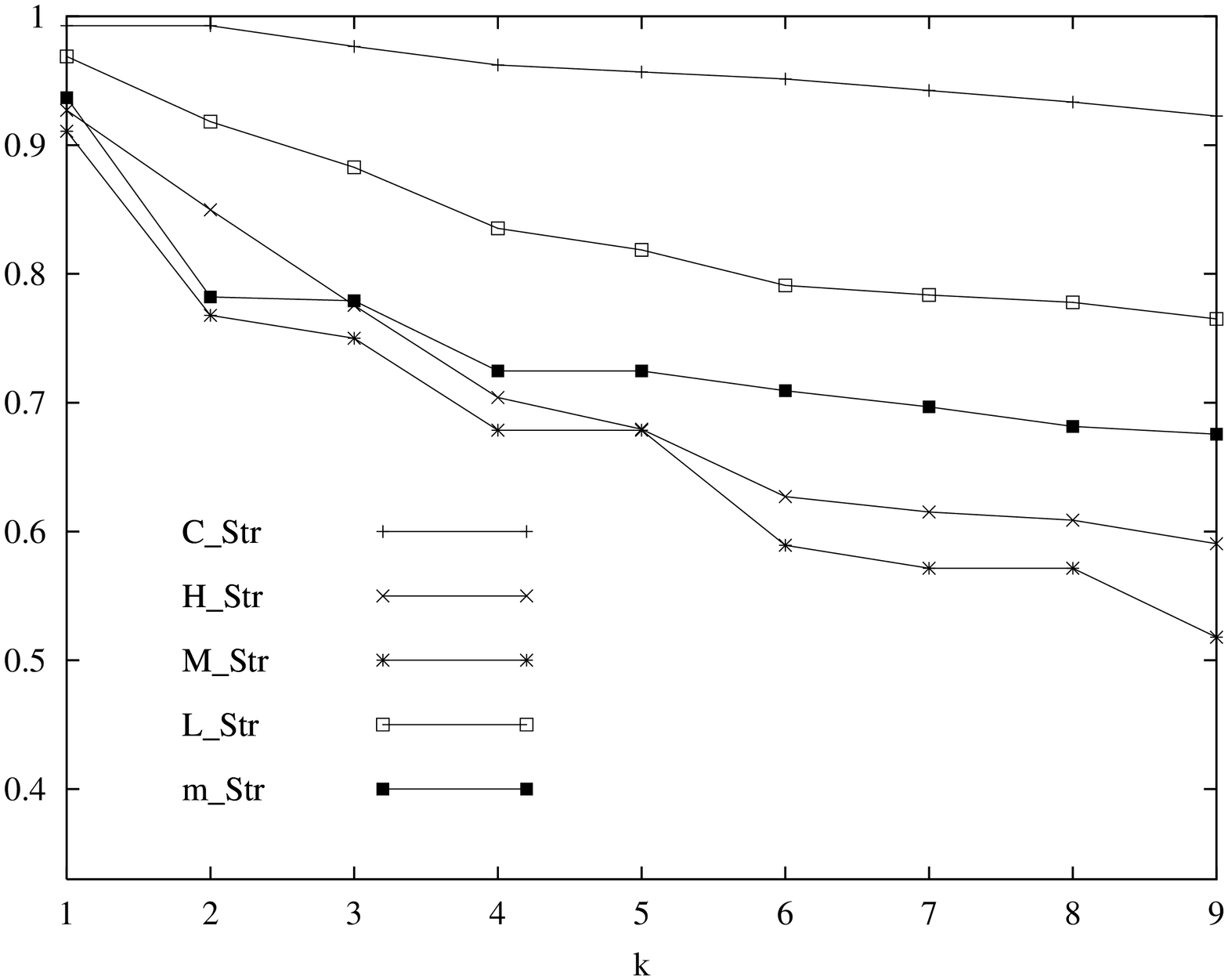}}
\subfigure[Random-50]{\label{PerfImp.50} \includegraphics[angle=-90, width=5.6cm,totalheight=3.6cm]{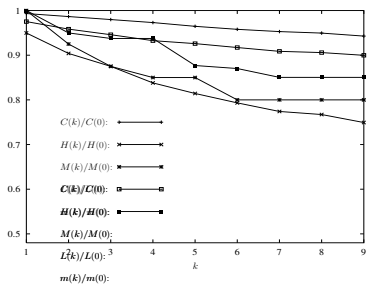}}
\subfigure[Random-60]{\label{PerfImp.60} \includegraphics[angle=-90, width=5.6cm,totalheight=3.6cm]{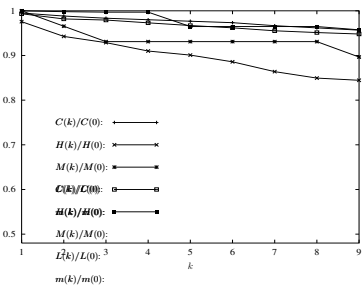}}
\caption{Performance improvement vs. the number of short-cut channels}
\label{Fig_PerfImp}
\end{figure}

In the above MIMO wireless environment, {\em Random-40} represents the instance with the random
node distribution and the radio range $40 m$, and Random-40 with $k$ short-cut channels is denoted
as {\em Random-40 (k)} $(k \ge 0)$. For instance Random-$40$ ($k$), the clustering coefficient,
characteristic path hop, maximum path hop,
characteristic path length, and the maximum path length are denoted as $C(k)$, $H(k)$, $M(k)$,
$L(k)$, and $m(k)$
$(k \ge 0)$ respectively. The power constant $\alpha$ is set to $1$, and similar results can be got
for the other value of $\alpha$. The topologies investigated in this work and some related information
are illustrated in Table \ref{Tbl_Topology}.

As in Table \ref{Tbl_Topology}, the clustering coefficient and path hop of the topologies without short-cut
channels, which are investigated in this work, are much higher than those of the random graph, except the
Grid-35 instance, where $C(0)=0$. This is due to the broadcasting nature of the wireless networks, which
greatly increase the number of a node's neighbors, which are also neighbors.

In the MIMO wireless network with one compound channel, we firstly limit the distance between the two
terminals of a short-cut by an upper bound of $RadiiRatio\times R_0=5\times R_0$, and
evaluate the performance as the number
of the short-cut channels $k$ changes \footnote{We mainly focus on the instances with random node distribution
in this work, since they are more practical. Similar conclusion
exists for the other instances.}.
Let $E_i$ be the set of edges over the $i^{th}$
short-cut channel, and $SC_i = E_1 \cup \cdots \cup E_i$, where $i \ge 1$.

The number of edges added over
each short-cut channel is shown in Table \ref{Tbl_SCEdge}. As in Fig. \ref{Fig_PerfImp}, an interesting
trend exists for all the instances. With only a few edges added (about $20 - 30$ edges for each short-cut channel),
the path hop is reduced drastically.
The characteristic path hop of the instance Random-60 with only $9$ short-cut channels is reduced by $16\%$, and
even better improvement can be got for the instances with smaller radio range, e.g. $25\%$ and $41\%$ for the
instances Random-50 and Random-40 respectively. Similar trends exist for the maximum path hop. Besides the
improvement of the topology-related topics, the power required by a request is also reduced greatly. As
illustrated in Fig. \ref{Fig_PerfImp}, the characteristic path lengths are reduced by $24\%$, $11\%$, and $6\%$
for Random-40, Random-50, and Random-60 respectively. The reduction curves of maximum path lengths are similar.

The above observations are different from those of the previous small-world networks, including Ahmed~Helmy's
virtual topology model on the wireless networks, in which the path hops can be greatly reduced with only a few
edges added, and further adding doesn't contribute much.
This is due to the construction scheme of
the short-cuts, which makes that there is no interference between any two short-cuts, or between a short-cut
and a normal edge. Table \ref{Tbl_SCEdge} and Fig. \ref{Fig_PerfImp} also suggest that by introducing only a
few short-cut channels into the MIMO wireless networks, the path hop and path length may be greatly reduced,
and these short-cut channels are especially beneficial for those instances with small radio range,
e.g. the wireless sensor networks \cite{Sensor}.

\psfrag{eStr}{\LARGE{$|SC_9|/|E_0|$: }}
\psfrag{CStr}{\LARGE{$C(9)/C(0)$: }}
\psfrag{HStr}{\LARGE{$H(9)/H(0)$: }}
\psfrag{MStr}{\LARGE{$M(9)/M(0)$: }}
\psfrag{LStr}{\LARGE{$L(9)/L(0)$: }}
\psfrag{mStr}{\LARGE{$m(9)/m(0)$: }}
\psfrag{RadiiRatio}{\LARGE{$RadiiRatio$}}

\begin{figure}
\centering
\subfigure[Random-40(9)]{\label{RadiiRatio.40} \includegraphics[angle=-90, width=5.6cm,totalheight=4cm]{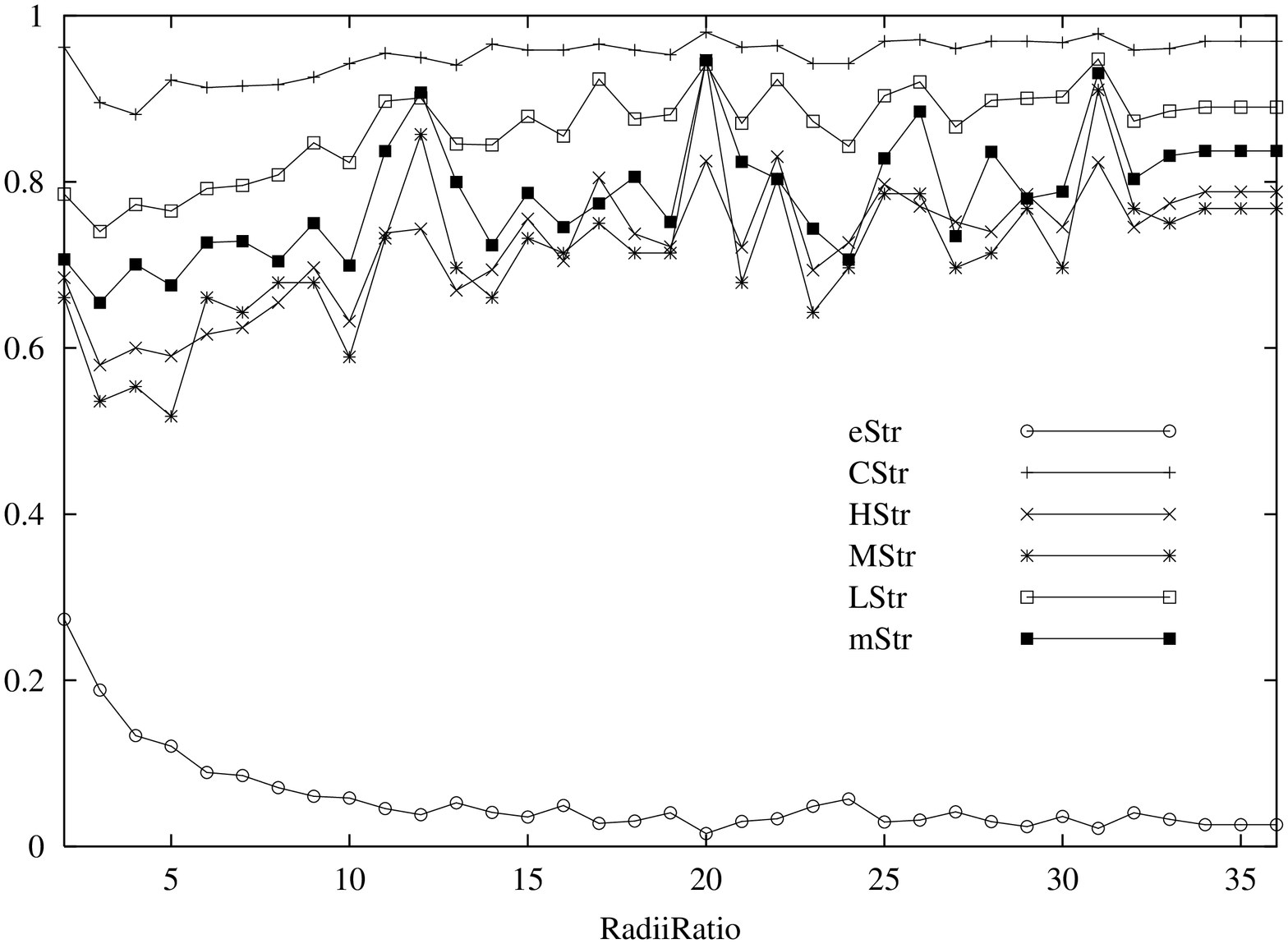}}
\subfigure[Random-50(9)]{\label{RadiiRatio.50} \includegraphics[angle=-90, width=5.6cm,totalheight=4cm]{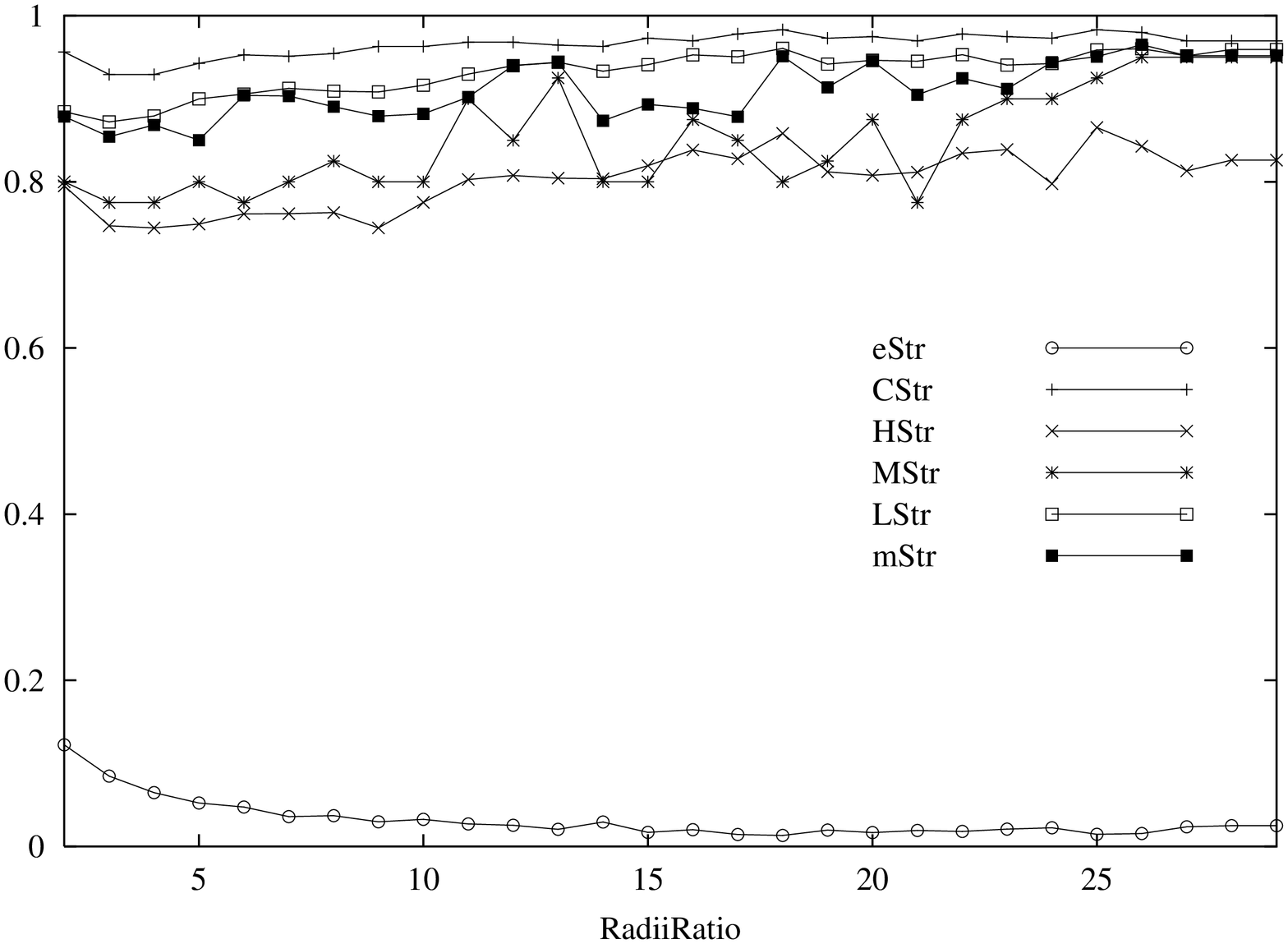}}
\subfigure[Random-60(9)]{\label{RadiiRatio.60} \includegraphics[angle=-90, width=5.6cm,totalheight=4cm]{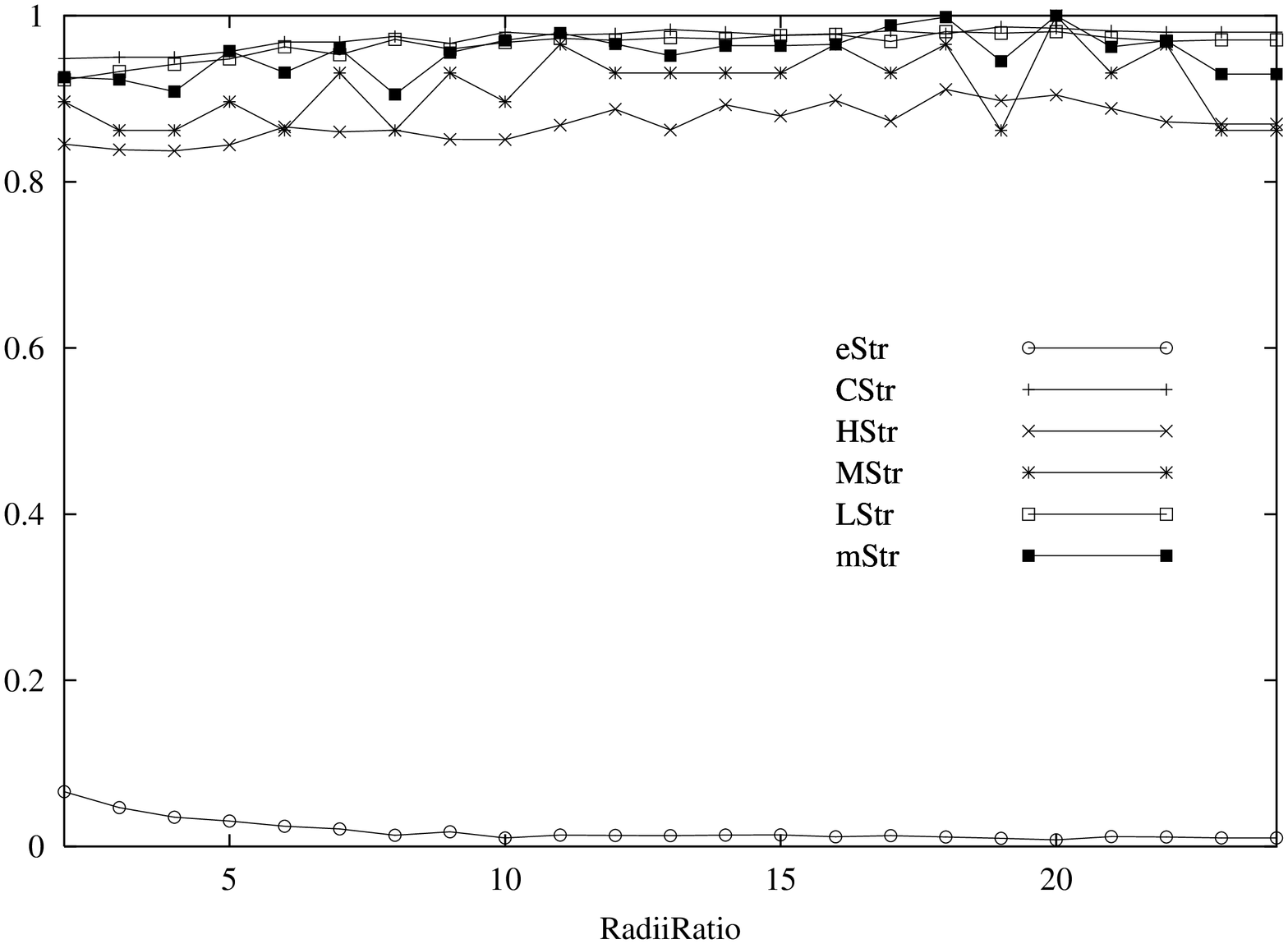}}
\caption{Performance improvement vs. $RadiiRatio$.}
\label{Fig_RadiiRatio}
\end{figure}

\begin{table}
\renewcommand{\arraystretch}{1.3}
\caption{Value of the curves at $RaiiRatio = 5$ vs. the minimum value.}
\label{Tbl_Value}
\centering
\begin{tabular}{|c|c|c|c|}
\hline
 & Random-40 & Random-50 & Random-60 \\
\hline
{\bf $SC'/\min{\{|SC_9|/|E_0|\}}$ } & 7.79 & 3.98 & 3.90 \\
\hline
{\bf $C'/\min{\{C(9)/C(0)\}}$ }     & 1.04 & 1.01 & 1.00  \\
\hline
{\bf $H'/\min{\{H(9)/H(0)\}}$ }     & 1.01 & 1.00 & 1.00 \\
\hline
{\bf $M'/\min{\{M(9)/M(0)\}}$ }     & 1.00 & 1.03 & 1.04        \\
\hline
{\bf $L'/\min{\{L(9)/L(0)\}}$ }     & 1.03 & 1.03 & 1.02 \\
\hline
{\bf $m'/\min{\{m(9)/m(0)\}}$ }     & 1.03 & 1.00 & 1.05 \\
\hline

\end{tabular}
\end{table}

Besides the number of the short-cut channels, we are also interested in how the upper bound of
short-cuts will reduce the evaluated topics. We conduct this set
of experiments on the three instances, Random-40(9), Random-50(9), and Random-60(9).
In the area of $1km \times 1km$, the distance between two nodes is at most $1000 \sqrt{2} m$. So we only
need to consider $RadiiRatio \le (1000 \sqrt{2}) / R_0 $, which are $36$, $29$, and $24$ for the three
instances respectively. Besides the topics considered in the above, we also study how the number of short-cuts
added is affected by $RadiiRatio$, which is denoted as $|SC_9|/|E_0|$.

As in Fig. \ref{Fig_RadiiRatio}, there exists an interesting observation for the three instances: all the
curves reach their minimum or near-minimum value at $RadiiRatio = 5$, except the curves of $|SC_9|/|E_0|$,
and there is almost no further contribution when $RadiiRatio > 5$.
Let the value of the curves at $RadiiRatio=5$ be $SC'$, $C'$, $H'$, $M'$, $L'$, and $m'$ respectively.
Table \ref{Tbl_Value} illustrates how close these values and the minimum ones are. The difference between
them are less than $5\%$ for all the topics except $SC'/\min{\{|SC_9|/|E_0|\}}$.
The ratio of the number of short-cuts $SC'/\min{\{|SC_9|/|E_0|\}}$ is a bit higher than the other
evaluated topics. But according to the
above discussion, the short-cuts can be constructed distributedly. So the high ratio
of the number of short-cuts will not be a great holdback for
the global performance of the MIMO wireless networks.

The above discussion suggests that by limiting the distance between the two terminals of a short-cut
with an upper bound $RadiiRatio \times R_0$ for a certain value of $RadiiRatio$, e.g. $RadiiRatio \sim 5$,
the clustering, path hop and path length of the MIMO wireless networks can be reduced to the minimum
or near-minimum value.

\section{Conclusions and Future Work}\label{Sec_5}

In this work, the small world concept is introduced into the physical topology of MIMO wireless networks.
By implementing several short-cut channels over the radio spectrum, the clustering, path hop and path
length of the wireless networks can be greatly reduced, especially for those networks with small
radio range. With the limited number of short-cut channels and the limited upper bound of the distance
between the two terminals of a short-cut, the small world structure can be easily constructed in a
distributed manner. Incorporating the small world structure into the practical wireless protocols will
be our future work.

\section*{Acknowledgements}

We would like to thank Prof. Ahmed~Helmy for his kindly discussion on the virtual small-world topology
in the wireless networks, and pointing out the two related papers \cite{small_world_7,small_world_8}.
We would also like to thank Dr. Yunnan~Wu for his valuable comments and pointing out the paper
\cite{MIMO}.


\end{document}